\documentclass[5p,twocolumn]{elsarticle}
\usepackage{amssymb}
\usepackage{epsf,graphicx}
\begin{document}
\title{Study of peculiarities of the thermal expansion of zirconium thin films 
by molecular-dynamics simulation}
\author{E.B. Dolgusheva and V.Yu. Trubitsin}
\date{}
\ead{elena@ftiudm.ru}
\address{Physical-Technical Institute,
Ural Branch of Russian Academy of Sciences,
132 Kirov Str.,426001 Izhevsk, Russia}
%\preprint{APS/123-QED}
\date{\today}

\begin{abstract}

The peculiarities of thermal expansion of $bcc$ and $fcc$ zirconium films with
(100) and (110) crystallographic orientations are studied at a constant zero
pressure by the molecular dynamics (MD) method with a many-body interatomic
interaction potential obtained in the embedded atom model. It is shown that
after relaxation the cubic lattices become tetragonal ($bct$ and $fct$), and for
the metastable $fct$ films the linear coefficients of thermal expansion in the
film plane have a negative value in a wide temperatuThe peculiarities of thermal expansion of $bcc$ and $fcc$ zirconium films with
(100) and (110) crystallographic orientations are studied at a constant zero
pressure by the molecular dynamics (MD) method with a many-body interatomic
interaction potential obtained in the embedded atom model. It is shown that
after relaxation the cubic lattices become tetragonal ($bct$ and $fct$), and for
the metastable $fct$ films the linear coefficients of thermal expansion in the
film plane have a negative value in a wide temperature range. The total and
local vibrational density of states (VDOS)  polarized along the $x,y,z$ axes
is calculated for the surface and interior layers of $bct$ and $fct$ Zr films
as a function of the temperature. It is shown that the peculiarities of the
behavior of the vibrational density of  states of surface atomic layers
manifest themselves in the anisotropy of the changes of the film lattice
parameters with temperature variation. A decrease in the lattice parameters with
increasing temperature is observed in the directions where there occurs a
softening of the local vibrational modes. re range. The total and
local vibrational density of states (VDOS)  polarized along the $x,y,z$ axes
is calculated for the surface and interior layers of $bct$ and $fct$ Zr films
as a function of the temperature. It is shown that the peculiarities of the
behavior of the vibrational density of  states of surface atomic layers
manifest themselves in the anisotropy of the changes of the film lattice
parameters with temperature variation. A decrease in the lattice parameters with
increasing temperature is observed in the directions where there occurs a
softening of the local vibrational modes.


\end{abstract}

\maketitle

\section { Introduction}

In recent years, an ever increasing number of papers have been devoted to the
study of so-called ``negative'' materials having negative coefficients of
thermal expansion  \cite{Science-96,Srikari,Masaaki,Michael}.
In 1996 the research team led by John Evans published the paper
\cite{Science-96} in which a negative coefficient of thermal expansion (CTE) 
was found for the $ZrW_{2}O_{8}$ ceramics in a wide range of temperatures from
$0.3K$ to $1050K$. Later many other compounds with similar properties, named
Negative Thermal Expansion (NTE) materials, have been discovered \cite{Lind}.

These materials with unusual properties are interesting not only theoretically,
but also from an industrial standpoint. Materials with controlled thermal
expansion can find several practical applications, such as switches and sensors,
while materials with zero thermal expansion in a wide temperature range will be
used in high-precision engineering. In instrument-making industry materials with
strictly regulated values of linear CTE in specified
temperature ranges of operation are in great demand.
Recently the Evans team  has created a material whose
properties can be controlled throughout the thermal expansion range: from
positive through zero to negative CTE \cite{Evans-2013}. The
composite oxide $Zr Mo_{2}O_{8}$ is a well studied compound distinguished by a
negative CTE, whereas the tin-containing analog of
this cubic material, $Sn Mo_{2}O_{8}$, expands when heated.
The authors of that paper state also to have demonstrated the possibility of
introducing various amounts of zirconium into the lattice so that the obtained
ceramic material can have a positive, negative or zero CTE. In
essence the ``positive'' nature of the tin-containing material is balanced by
``negative'' zirconium, both metals being key elements of the same crystalline
structure.
Additionally, in Refs.\cite{Mittal,Erst} it was shown by inelastic neutron
scattering that in NTE structures a softening is observed in the VDOS
curve. Thus a negative thermal expansion
coefficient is observed in materials which are in a metastable state 
(for example, near a structural transition), have a strongly anharmonic
character  of interatomic interaction, and/or a softening of the phonon modes is
evidenced in the VDOS curve, which results in a negative value
of the Gr\"{u}neisen parameter.

Zirconium is known to be a strongly anharmonic metal. As we see, it plays a
decisive role in the formation of negative CTE in the NTE structures. We
believe, then, that the study of the thermal expansion of Zr nanofilms is an
important task, because elucidation of the dependence of linear CTE on the
temperature and film thickness will make it possible to reveal the peculiarities
of the formation of thermodynamical properties of nanofilms and draw some
conclusions about the influence of the surface structure on their properties. 
Clarification of the nature of special physical features of nanosystems is of
interest for both fundamental problems of the condensed matter physics and
possible practical applications. 

Earlier \cite{our-2013}, we constructed a phase diagram Temperature-Thickness
for zirconium films. 
This diagram presents the regions of final structures
stable for no less than 1 ns, one time step being equal to 1 fs. Here we should
mention some important points. Firstly, the size effect was detected. Two
characteristic sizes may be noted. The first one is of 5 unit cells ($\sim2nm$):
up to this size the film just falls apart, and the second critical size is the
film thickness of $17$ unit cells ($\sim6.1 nm$) at which the mechanism and
sequence of
structural transformations change: from this size on the $bcc$ (001) film
remains stable in a sertain temperature range. Secondly, what is especially
important, structures not observed in pure Zr in the usual bulk state were
found. These are $fcc$ and orthorhombic structures. Recall that at atmospheric
pressure only two phases are observed in Zr bulk samples: a high-temperature
$bcc$ phase, and an $hcp$ phase found below $~1136 K$. Thus, zirconium provides
a possibility of studying the properties of a strongly anharmonic material
existing in both stable and metastable states, in conditions close to structural
transformations.

The molecular dynamics method makes it possible to determine the linear
CTE in different crystallographic directions, namely, in the
direction normal to the film surface, and in the plane of the film atomic
layers, which allows the study of the film thermal expansion anisotropy.
Furthermore, this method enables the investigation of physical characteristics,
such as the lattice dynamics, separately for the surface and interior atomic
layers of the film. In this work the MD method is used to study the
peculiarities of thermal expansion of the zirconium thin films with cubic ($bcc,
fcc$) crystal structures for different crystallographic orientations of the
surface at a constant zero pressure. We also performed a comparison of the
temperature behavior of the lattice parameters and vibrational density of
states of surface and interior atomic layers, polarized both in the film plane
and in the direction normal to the film surface.

\section{Calculation method}

\begin{figure}[tbh]
\begin{center}
\resizebox{0.92\columnwidth}{!}{\includegraphics*[angle=-90]{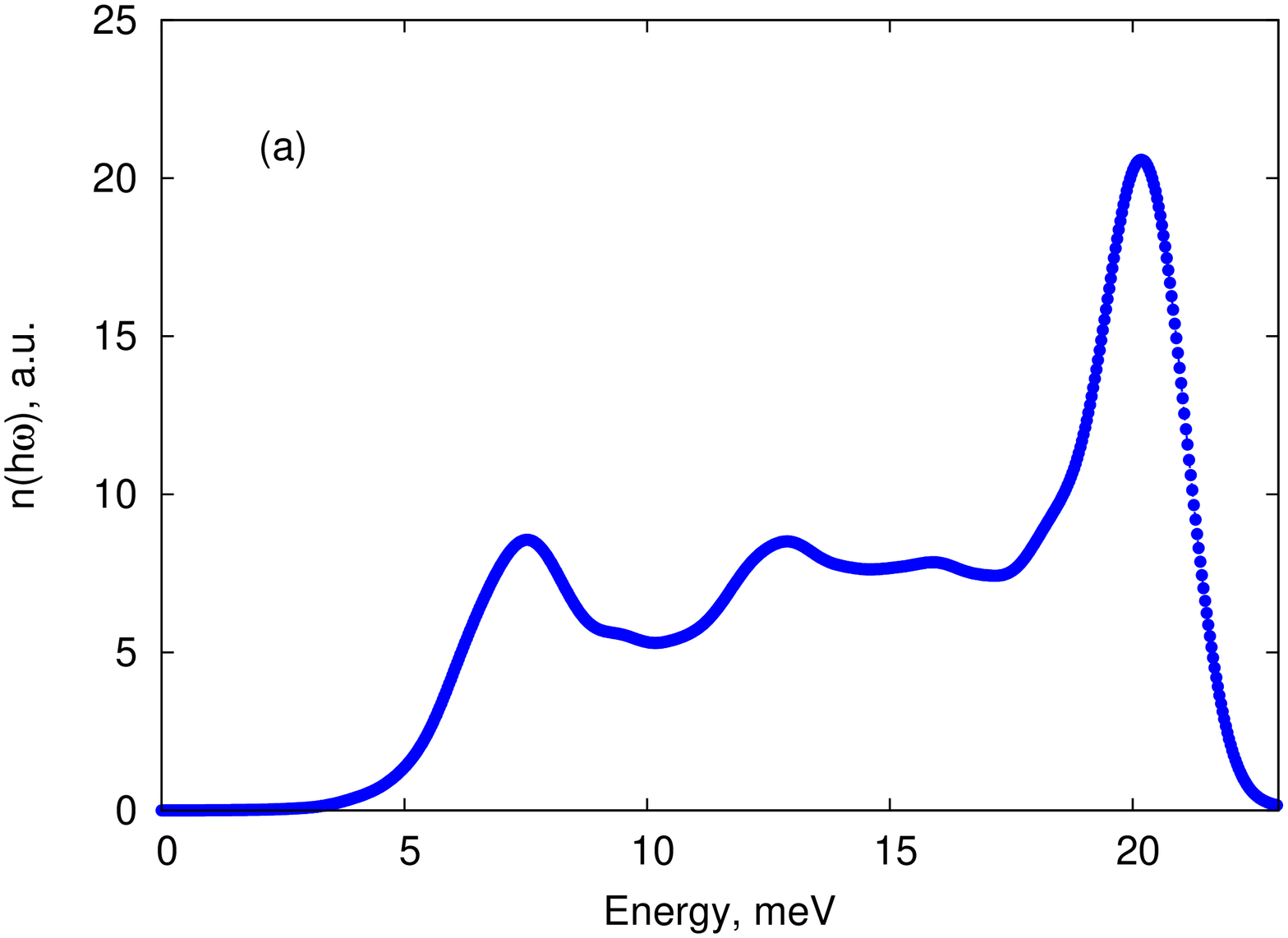}}
\resizebox{0.95\columnwidth}{!}{\includegraphics*[angle=0]{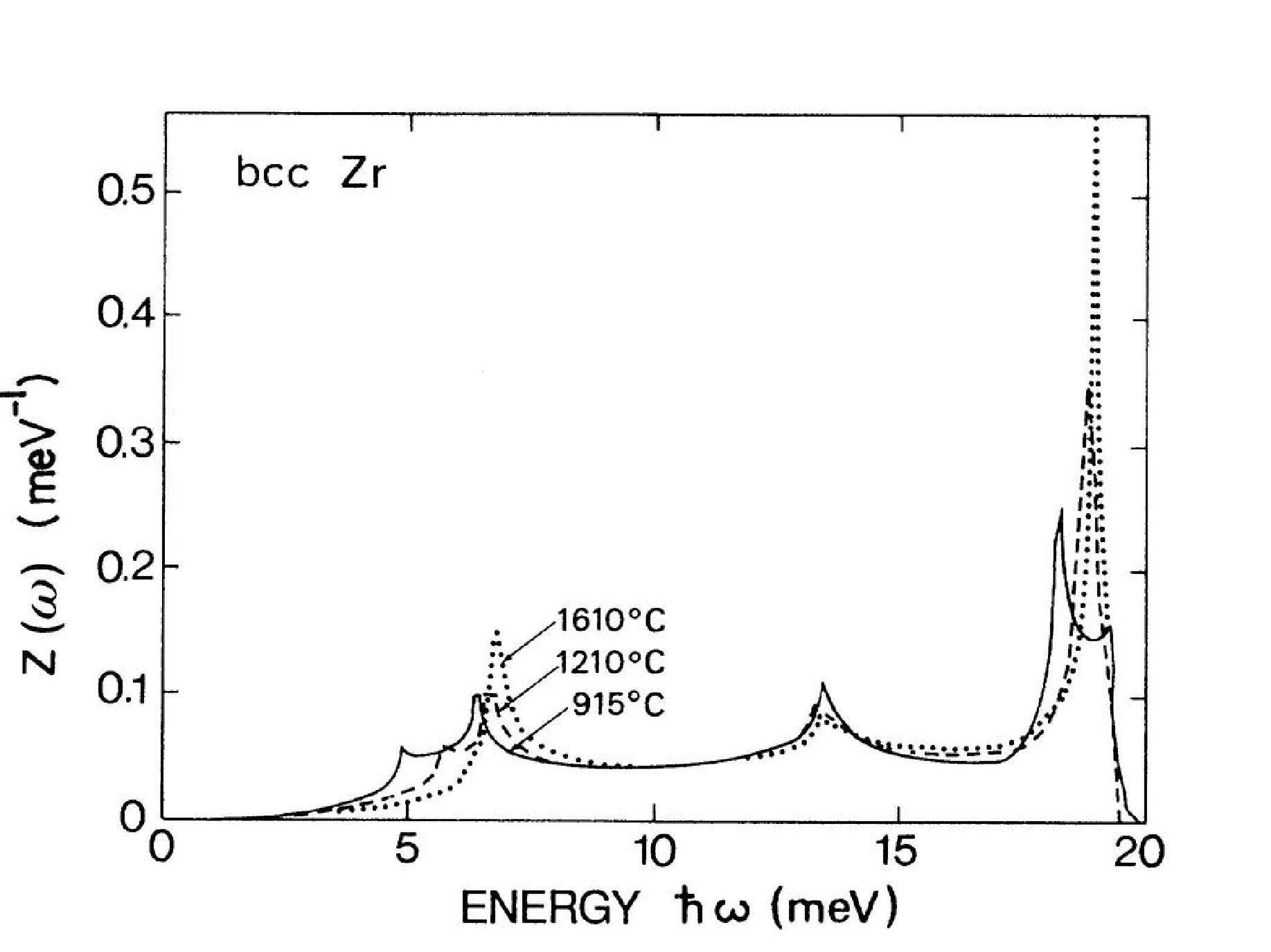}}
\caption{Vibrational density of  states of bcc Zr. (a) calculated for a
crystallite with cyclic boundary conditions at T = 900 K; (b) experimental
values \cite{Phonon-bcc}.}

\label{Fig1}
\end{center}
\end{figure}

The physical characteristics of zirconium films were studied by  the MD method
using the standard XMD package \cite{J-Rifkin}. The many-body potential (\#2)
from Ref.\cite{Mendelev-Ack} constructed within the embedded atom model
\cite{Daw-Baskes} was chosen to describe the interatomic interaction in
zirconium. In Ref.\cite{Mendelev-Ack} it was shown that this potential allows
one to obtain, to a high degree of accuracy, the $bcc$ and $hcp$ lattice
parameters of zirconium, cohesive energy, elastic constants, melting
temperature, and other physical characteristics. The bulk phonon dispersion
curves calculated with this potential along the symmetrical directions of the
Brillouin zone of $bcc$ zirconium at different temperatures were obtained in our
paper \cite{our-2009}. A comparison with the experimental data shows that the
potential chosen allows one to reproduce the experimentally observed features of
the Zr phonon spectrum including the softening of the transverse N-phonon
with decreasing temperature, and so it may be successfully used in calculating
vibrational and thermal characteristics of zirconium films.

In this paper we present only the vibrational density of states for bulk
zirconium (see Fig.\ref{Fig1}) as compared with the experiment
\cite{Phonon-bcc}. The total VDOS, g($\omega$), was calculated as the Fourier
transform of the autocorrelation velocity function, the velocity was averaged
over all atoms on an interval of 10 ps after a relaxation period of 100 ps at a
given temperature:

\begin{equation}
g(w)={\int{dt\frac{\sum_{i=1}^{N}\langle
{\bf v_{i}}(t)|{\bf v_{i}}(0)\rangle}{\sum_{i=1}^{N}\langle
{\bf v_{i}}(0)|{\bf v_{i}}(0)\rangle}\exp(iwt).}}
\end{equation}

Here ${\bf v_{i}}(t)$ and ${\bf v_{i}}(0)$ are the velocities of the $i$-th atom
at time $t$ and the initial time, respectively. $\textless$  $\textgreater$ 
denotes the averaging over time and the sum is taken over all atoms.
For the calculation of the local VDOS (for example, the surface atoms VDOS) in
the sum were considered only the velocities of  the surface layer atoms. The
local VDOS then  were multiplied by the weight coefficient.
To obtain the polarized VDOS all the velocities were multiplied by the
polarization vector $\bf p$ in calculating the autocorrelation function:

\begin{equation}
{\bf v_{p}}(t)={\sum_{i=1}^{N}{\bf p}} {\cdot{\bf v_{i}}(t)}.
\end{equation}

As seen from Fig.\ref{Fig1}, the positions of the main peaks are in good
agreement with the experiment.  

In calculating the lattice parameters, the coordinates of all the crystallite
atoms were first averaged over a time interval of 50 ps, then the average values
were taken over the atomic layer, thereafter the distance between the atomic
layers in the given direction $(x, y, z)$ was found. 

Note that in calculating the interlayer distance along the $z$ axis the
near-surface layers  (5 layers below the surface on each side) were excluded
from consideration because of the fact that
after relaxation near the film surface significant deviations from the
interlayer distance values in the film interior are observed. 

The coefficients of thermal expansion were calculated as follows: 

\begin{equation}
\beta=\frac{1}{V} \frac{dV}{dT},
\end{equation}

where $\beta$ is the film CTE; $V$ is the average volume per atom
only in the film interior at the initial temperature;
$dV$ is the relative change of this volume on heating the crystallite by $dT$
degrees. For the calculation of the linear CTE from our simulation we used the
following expression:

\begin{equation}
\alpha_{x,y,z}=\frac{1}{L_{X,Y,Z}} \frac{dL_{X,Y,Z}}{dT}.
\end{equation}

Here $\alpha_{z}$ is the linear CTE in the direction normal to the film plane
along the $z$ axis; $\alpha_{x,y}$ are the linear CTE in the film plane along
the $x$ and $y$ axes; $L$ is the average distance between the atomic  layers at
the initial temperature along the direction chosen; $dL$ is the relative change
of this distance on heating the crystallite by $dT$ degrees.

\section{Lattice Parameters and Vibrational Density of States}

Note at once that the lattice cubic symmetry in $bcc$ and $fcc$ films changes
after relaxation to tetragonal. As a result, the $bcc$ films turn into $bct$,
and $fcc$ into $fct$ ones.

\begin{figure}[tbh]
\begin{center}
\resizebox{0.99\columnwidth}{!}{\includegraphics*[angle=0]{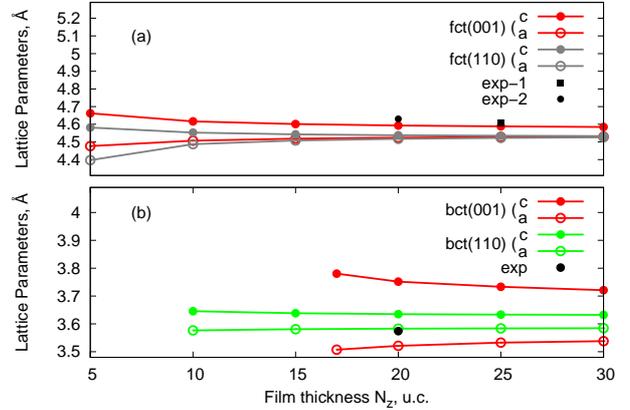}}
\caption
{Lattice parameters $a$ and $c$ for the films after relaxation: (a) for $fcc$
films: (b) for $bcc$ films. The experimental values for $fcc$ Zr film  were
obtained:  in \cite{Chopra,Ji} (exp-1); in \cite{Hill} (exp-2), and in
\cite{Tonkov} for a bulk $bcc$ Zr sample.}

\label{Fig2}
\end{center}
\end{figure}

Figure \ref{Fig2} presents the lattice parameters $a$ and $c$ averaged over 50
ps for films of different thickness indicated on the abscissa in unit cells
(u.c.) of the corresponding structures. For $fct$ films the calculation was
performed at $T = 300 K$ (Fig.\ref{Fig2}(a)), and for $bct$ films at $T = 900K$
(Fig.\ref{Fig2}(b)). The statistical error was $0.001$ \AA $ $ at $300K$ and
$0.002$ \AA{}  at $900K$. The lattice 
parameters for $fct$ films with (001) and (110) crystallographic  orientations
are rather close in magnitude, while for $bct$ (001) and (110) films they differ
significantly. The $bct$ (001)  film is stabilized only from a thickness of 17
unit cells ($N_{z}=17u.c.$). In the same figure are shown the experimental
lattice parameters obtained in Refs. \cite{Chopra,Ji} (4.61\AA $ $ indicated by
black square), and  Ref.\cite{Hill} (4.63\AA  $ $ indicated by black point) for
$fcc$ films, and in Ref.\cite{Tonkov} (3.609\AA $ $) for a bulk $bcc$ zirconium
structure (fig.\ref{Fig2}(b)). The tetragonal distortion parameter, $c/a$,
decreases with increasing film thickness.
For $fct$ films it varies from $1.042$ to $1.011$. The tetragonality of $fct$
zirconium films was also observed experimentally \cite{Ji} with a tetragonal
distortion parameter $c/a$ equal to $1.0220 \pm 0.022.$

The thickness has a determining effect on various properties of the films, which
is due to their structural peculiarities and the arising mechanical stresses. We
calculated the averaged distances between atomic layers in films of different
thickness along the direction normal to the film plane, i.e. from one free
surface to the other. The size of the basic crystallite along the $x$ and $y$
axes  in all cases was equal to 24 u.c.
The $bct$ films were considered at $T=900K$, and the $fct$ ones at $T=300K$.

As shown by the calculations, significant variations in the average distance
between
atomic layers are observed in the interface regions of all films after
relaxation.
These areas with strong distortions were eliminated from further calculations
of the lattice parameters and, accordingly, of the coefficients of thermal
expansion.
In all the cases considered, five interlayer intervals were assigned to the
interface region. In calculating the lattice parameters under changes of
temperature, only the values obtained for the film interiors were taken into
consideration. In this way the lattice parameters of $bct$ and $fct$ films with
crystallographic orientations (001) and (110) were calculated at different
temperatures. At the same time, we calculated and analyzed the
distribution of the total and local  vibrational densities of  states and their
temperature dependences in order to determine the extent to which the structural
stability of $bct$ and $fct$ Zr films is affected by the interior and surface
vibrations. To this end the surface atomic layers were separated out and all
calculations were performed separately for surface and interior atomic layers.

\begin{figure}[tbh]
\begin{center}
\resizebox{0.99\columnwidth}{!}{\includegraphics*[angle=-90]{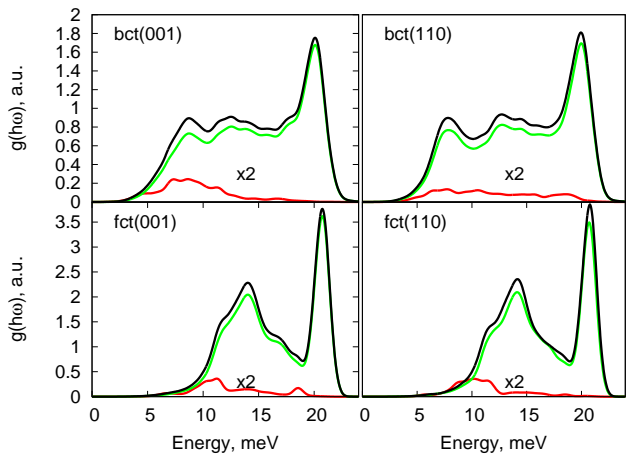}}
\caption{ Vibrational density of states:  for $bct$ (001) and (110)
films at $T=1200K$ (top panels); for $fct$ (001) and (110) films at
$T=300K$ (bottom panels). All the films are 20 u.c. thick. The contribution of
vibrational states from the surface atomic layers (multiplied by 2) is indicated
by the red line, from the interior atomic layers by the green line; the black
line shows the vibrational states of all the atoms of the film.} 
\label{Fig3}
\end{center}
\end{figure}

The top panels of Fig.\ref{Fig3} present the vibrational density of  states
for $bct$ films with (001) and (110) surface orientation, and the bottom
panels for $fct$ films with the same surface orientation. As can be seen,
for the  $bct$ (001) film all surface vibrations marked by the red line  are
in the low-frequency region, which is indicative of the dynamic instability of
the surface, while for the film with (110) surface these vibrations are
uniformly blurred  throughout the frequency range pointing to the dynamic
stability of the surface. In both cases VDOS for the whole film are practically
identical.

Returning to
the phase diagram Temperature-Thickness \cite{our-2013}, we can assume that the
critical size of 17 u.c (6.1nm) is due to the fact that with an increase of the
film thickness the proportion of surface atoms decreases and so does the
contribution from the surface vibration modes. As a result  the film with
(001) surface becomes energetically more favorable, whereas calculations of the
$bcc$ lattice energy at $T = 0K$ showed the (110) film to be energetically
favorable for all the film thicknesses (see Fig.6 in Ref.\cite{our-2013}). 

In the  $fct$(001) film surface vibrations give a small contribution to the
high-frequency region too, while in the  $fct$(110) film all surface vibration
modes are concentrated in the low-energy region. Our calculations on
heating the films show the  $fct$(001) film to be stable over a wider
temperature range: it undergoes structural transformation only at $1050 K$,
whereas the  $fct$(110) film even at $900 K$.

\begin{figure}[tbh]
\begin{center}
\resizebox{0.98\columnwidth}{!}{\includegraphics*[angle=-90]{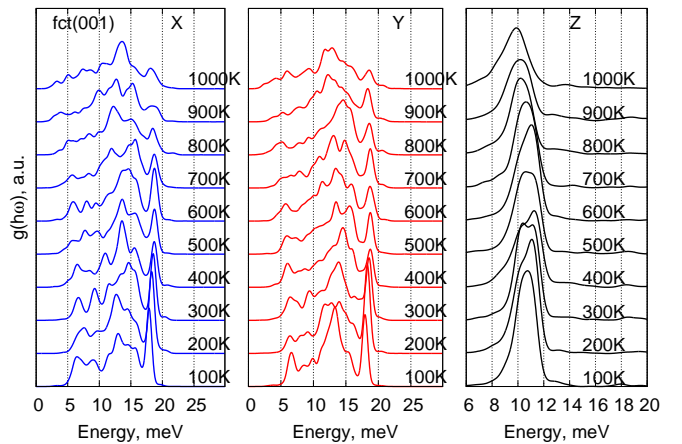}}
\caption{Local VDOS for the surface layers of the  $fct$(001) Zr film
polarized along the $x,y,z$ axes at different temperatures.}
\label{Fig4}
\end{center}
\end{figure}

Figure \ref{Fig4} shows the results of calculation of the local polarized VDOS 
for surface atoms of a  $fct$(001) film of thickness $N_{z}=20 u.c.$ as a
function of the temperature indicated on the right. Here one can see noticeable
shifts of the low-frequency peaks from the energy value of $7 meV$ to  
$2.5 meV$ for local VDOS along both $x$ and $y$ axes. At the same time the
contribution of high-frequency (about $18 meV$) vibrations polarized along these
axes substantially decreases. The position of the peak corresponding to the
surface atom displacements along the $z$ axis on heating first shifts slightly
to the region of higher energies, and starting from the temperature $T = 700 K$
this peak shifts in the opposite direction. This behavior, in our opinion, is
due to changes in interlayer distances in the interface region. At low
temperatures oscillations of the interlayer distances are observed, whereas with
increasing temperature these oscillations disappear, and all distances between
the atomic layers in the interface region (5 atomic layers from the surface)
increase. Just such a reconstruction occurs around $700 K$.  Before the
structural transformation (at $T = 1000 K$) all the atomic layers in the
near-surface area are spaced apart at larger intervals than the layers of the
film interior.

\begin{figure}[tbh]
\begin{center}
\resizebox{0.99\columnwidth}{!}{\includegraphics*[angle=-90]{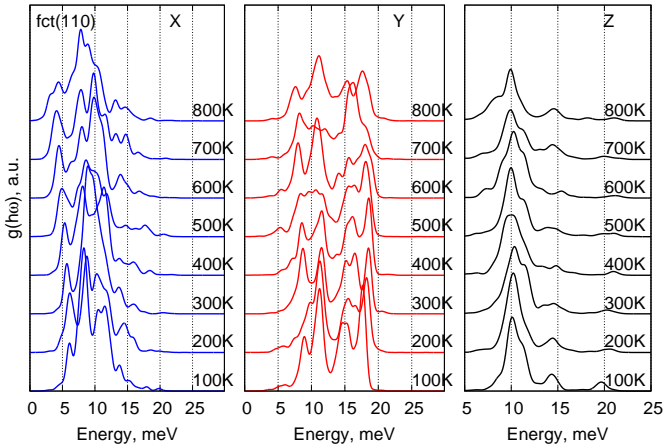}}
\caption{Local VDOS for the surface layers of the $fct$(110) Zr film
polarized along the $x,y,z$ axes at different temperatures.}
\label{Fig5}
\end{center}
\end{figure}

The  results of the local polarized VDOS calculation for surface atoms of a
$fct$ (110) film are presented in Fig.\ref{Fig5}: here the shift of the
low-frequency peaks is clearly observed only in one direction along the $x$
axis.
Note that different nature of the atomic layer displacement in the $fct$ (110)
and (001) films leads to the formation of different structures. The $fct$ (110)
film transforms into an $hcp$ (0001) at $900 K$, and the $fct$ (001) film into a
$bcc$ (110) one at a temperature of $1050 K$. Note also that on heating the
oscillations of interlayer distances persist in the $fct$ (110) film up to the
temperature of structural transition. 

\begin{figure}[tbh]
\begin{center}
\resizebox{0.99\columnwidth}{!}{\includegraphics*[angle=-90]{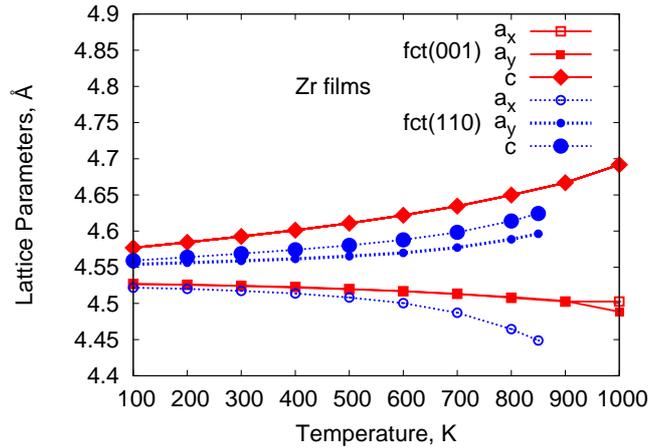}}
\caption{Temperature dependence of the lattice parameters of $fct$ Zr
films of thickness $20 u.c.$: red line for $fct$(001); blue line for
$fct$(110).}
\label{Fig6}
\end{center}
\end{figure}

We also calculated the temperature dependences of the lattice parameters $a_{x},
a_{y}$, and $c$ of $fct$ films 20 u.c. thick (see Fig.\ref{Fig6}). Red color
shows the
changes in the $fct$(001) lattice parameters. One can see that the parameter $c$
increases with temperature, while $a_{x}$ and $a_{y}$ decrease equally in the
whole temperature range considered. In the $fct$(110) film (blue color) only
the parameter $a_{x}$ decreases, whereas $a_{y}$ and $c$ grow.

Also were calculated the temperature dependences of the local VDOS polarized
along the $x, y$, and $z$ axes for both all interior atomic layers of $fct$
films and only two interior layers. However, no shift of vibrational modes to
the low-energy region was detected in either case. Thus, for metastable films
with $fct$ structure there is a clear correlation between the temperature
behavior of the surface atom vibrations and the lattice parameters.

Analogous VDOS calculations were performed for $bct$ zirconium films of
thickness
$N_{z}=20u.c.$ with (001) and (110) surface orientation. In the  $bct$(110)
film, on heating, all the lattice parameters $a_{x}, a_{y}, c$ increase, and no
softening of vibrational modes is observed  in the temperature behavior
of the local phonon densities polarized along the $x, y, z$ directions for both
surface and interior atomic layers. In the  $bct$(001) film, when heated from
$600 K$ to $1200 K$, all the lattice parameters also increase, but on further
heating to $1500 K$ (i.e. as the boundary of the structural reorientation
transition is approached) a slight decrease in the $a_{x}$ and $a_{y}$
parameters is observed, though the parameter $c$ continues growing. As seen from
the phase diagram Temperature-Thickness \cite{our-2013}, the $bct$ (001) film of
thickness $N_{z} = 20 u.c.$ is stable in the temperature range from $600 K$ to
$1500 K$, and at temperatures above $1500 K$ it undergoes an orientational
transformation into $bct$ (110). The decrease in the $a_{x}, a_{y}$ lattice
parameters of the $bct$(001) film in the range $1200-1500 K$ may be associated
with
stresses that arise in these directions resulting in a structural
transformation.

\section{Coefficients of thermal expansion }

\begin{figure}[tbh]
\begin{center}
\resizebox{0.49\columnwidth}{!}{\includegraphics*[angle=-90]{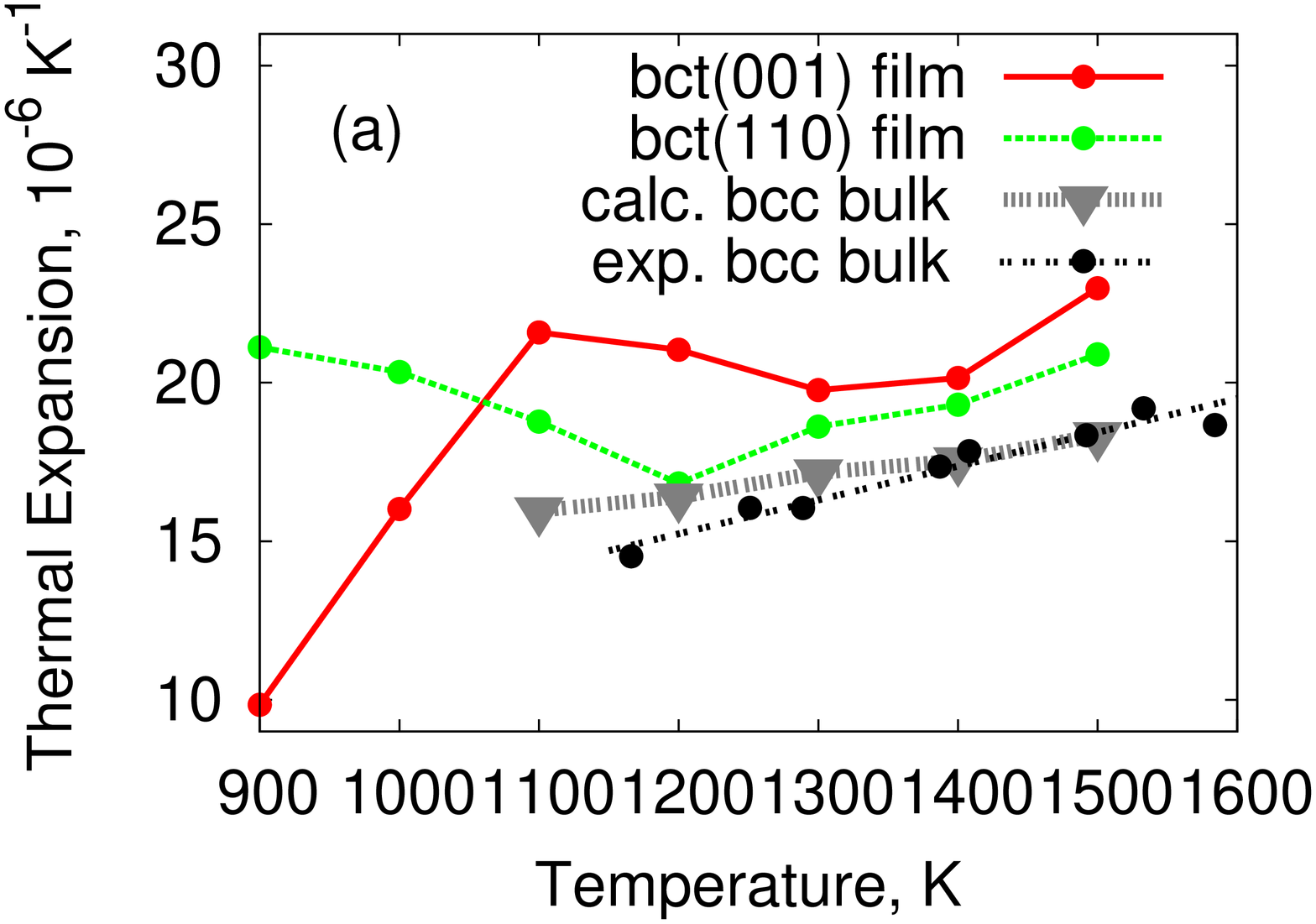}}
\resizebox{0.49\columnwidth}{!}{\includegraphics*[angle=-90]{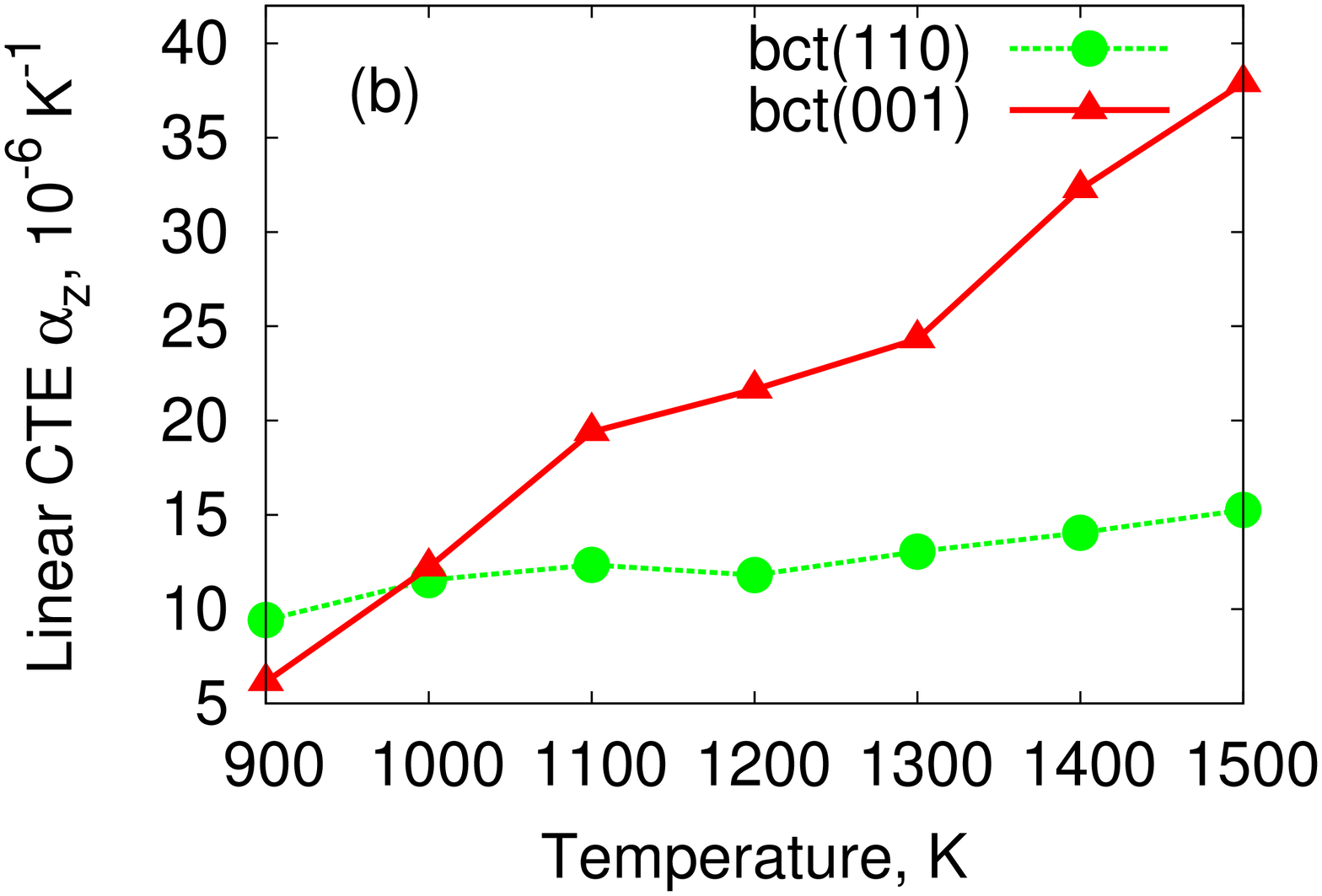}}
\resizebox{0.49\columnwidth}{!}{\includegraphics*[angle=-90]{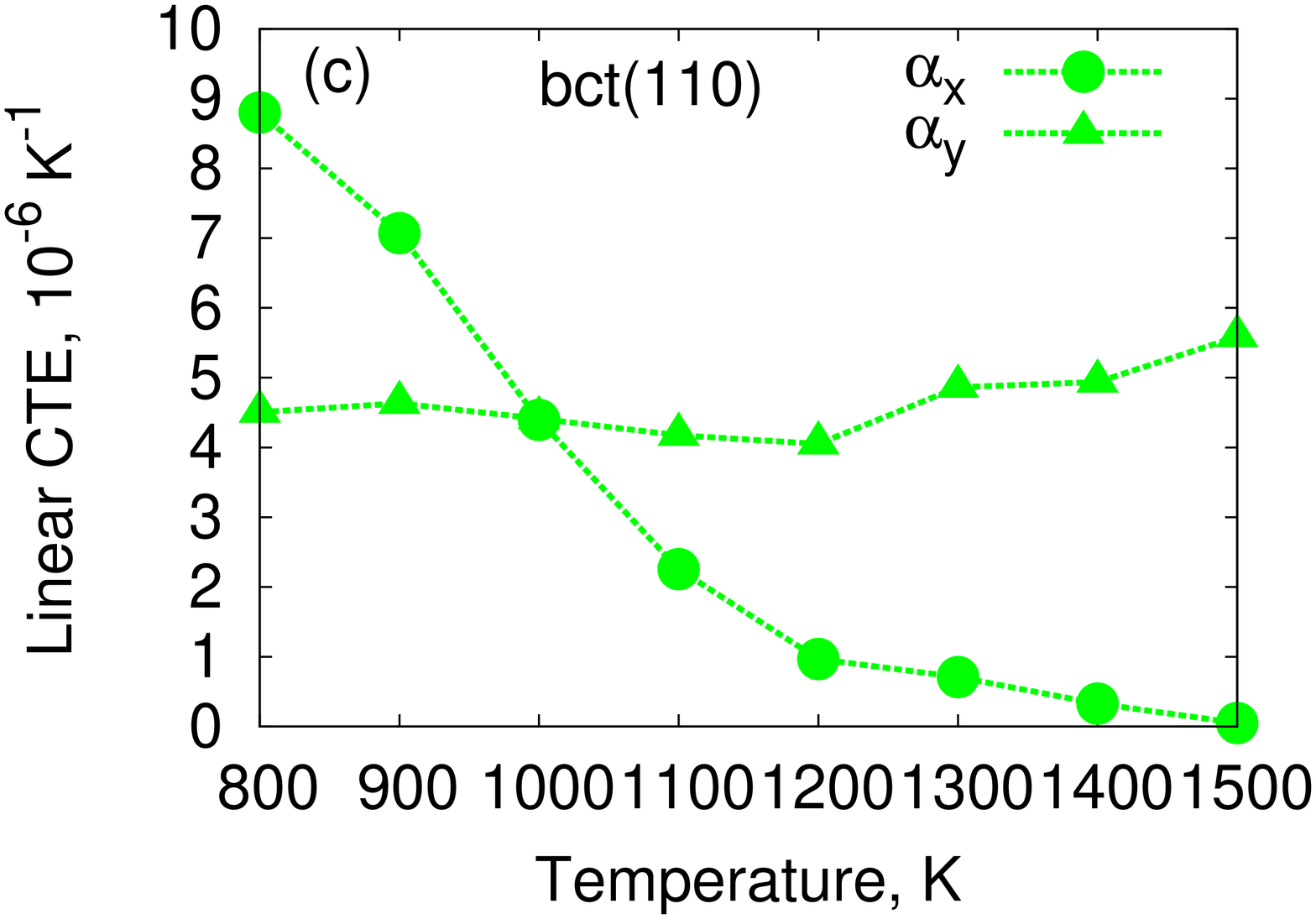}}
\resizebox{0.49\columnwidth}{!}{\includegraphics*[angle=-90]{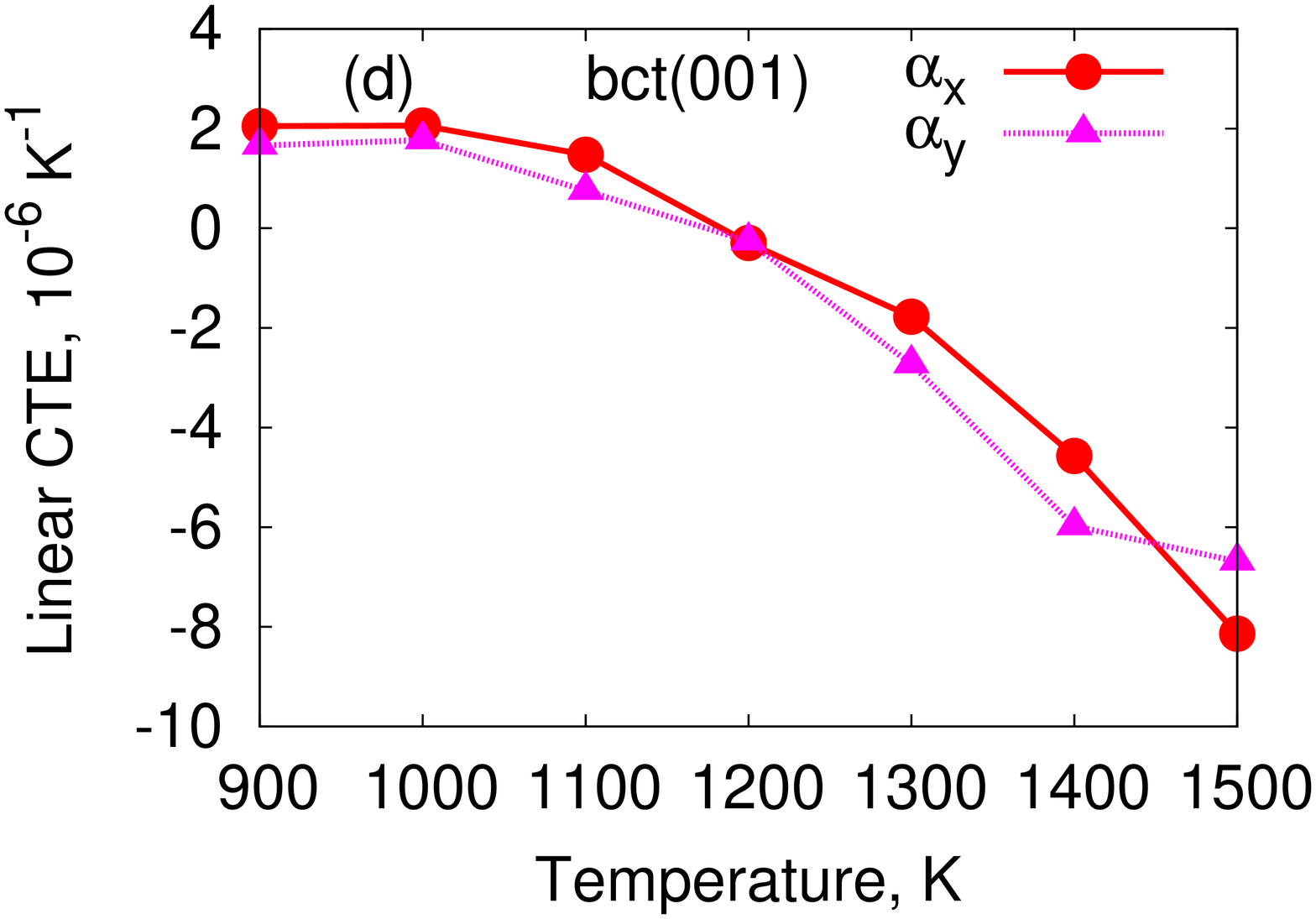}}
\caption{Coefficients of thermal expansion of Zr films as a function of
the temperature: (a) for $bct$ films (001) and (110) (red and green,
respectively), for a crystallite with cyclic boundary conditions (grey), and the
experimental data for the bulk sample\cite{Gordon.B.Skinner} (black); 
(b) linear CTE along the $z$ axis for the $bct$ (110)(red) and (110)(green)
films. Linear CTE along the $x$ and $y$ axes for the films $bct$ (110) (c), and
$bct$ (001) (d).}
\label{Fig7}
\end{center}
\end{figure}

As mentioned above, in the $bct$ (001) film the parameters $a_{x}$  and $a_{y}$
first increase and then (from $1200 K$ on) decrease. The appearance of such
two-dimensional stresses in the $bct$ (001) film with an increase in temperature
results in structural instability.
It is obvious that such changes in the lattice parameters result in nonlinear
variation the coefficients of thermal expansion  for films, as opposed to bulk
samples in which these coefficients vary linearly. In Fig.\ref{Fig7}(a)  are
presented the CTE for the zirconium films $bct$ (001) (red) and $bct$ (110)
(green). The figure also shows the CTE calculated for a $bcc$ crystallite with
cyclic boundary conditions (grey) and the experimental data for a bulk Zr
sample from Ref. \cite{Gordon.B.Skinner} (black). As seen from the
figure, the calculated and experimental CTE values for the ``bulk'' case are in
good agreement.

The linear coefficients of thermal expansion for $bct$ (110) and (001)
films are presented in Fig.\ref{Fig7}: $\alpha_{z}$ along the direction normal
to the film plane (b); $\alpha_{x,y}$ for the $bct$ (110) film (c); the same
coefficients for the $bct$ (001) film (d).

\begin{figure}[tbh]
\begin{center}
\resizebox{0.49\columnwidth}{!}{\includegraphics*[angle=-90]{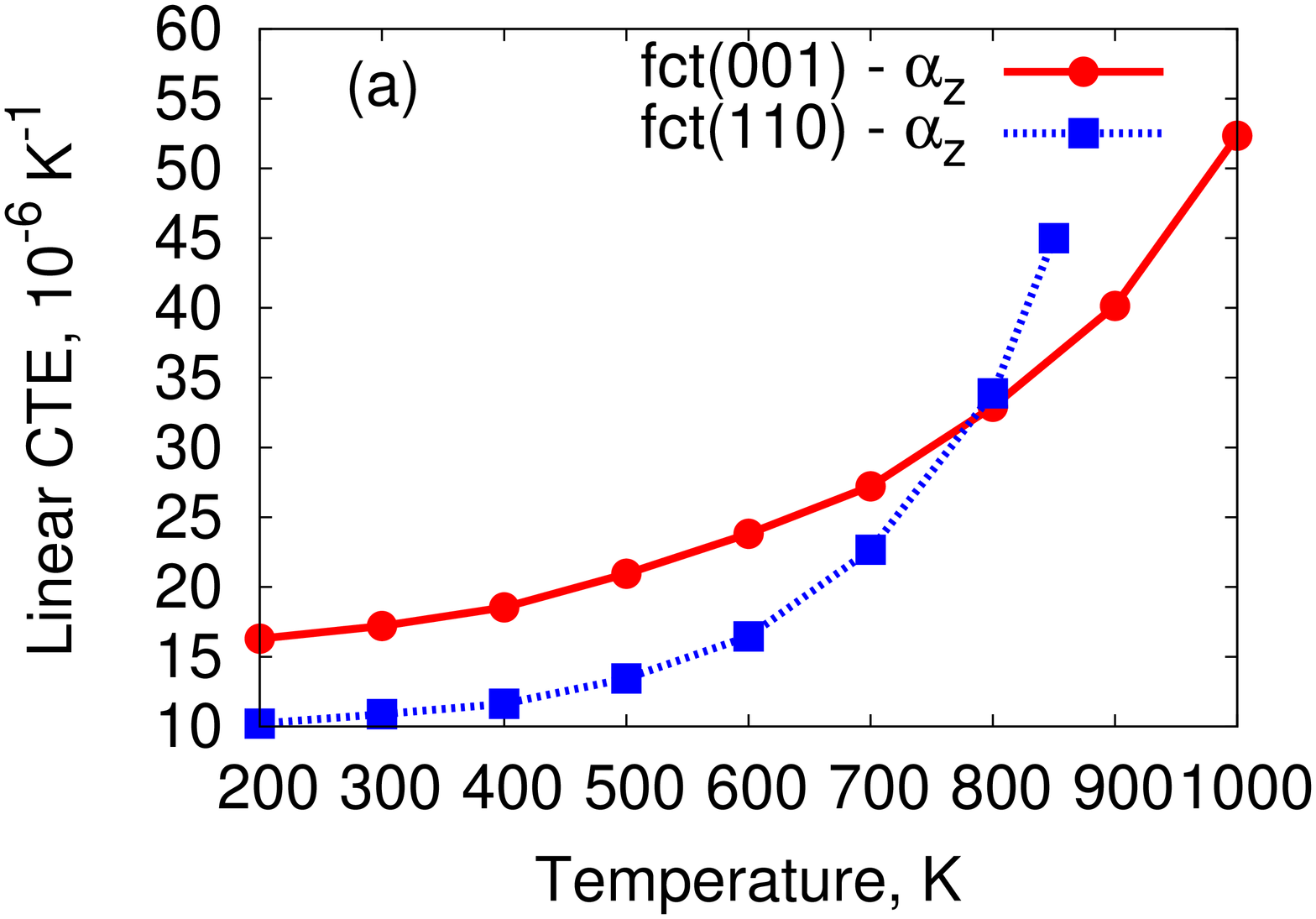}}
\resizebox{0.49\columnwidth}{!}{\includegraphics*[angle=-90]{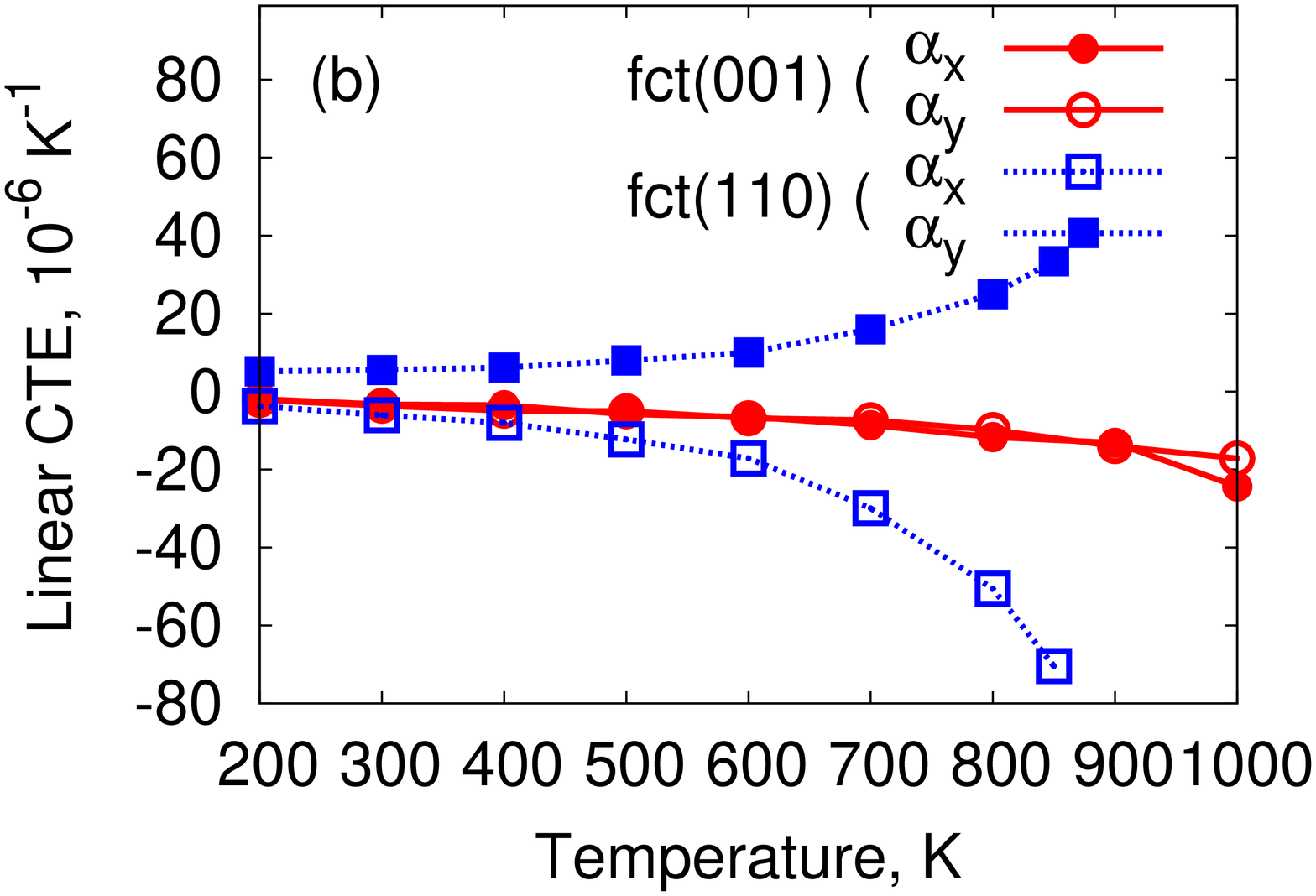}}
\resizebox{0.49\columnwidth}{!}{\includegraphics*[angle=-90]{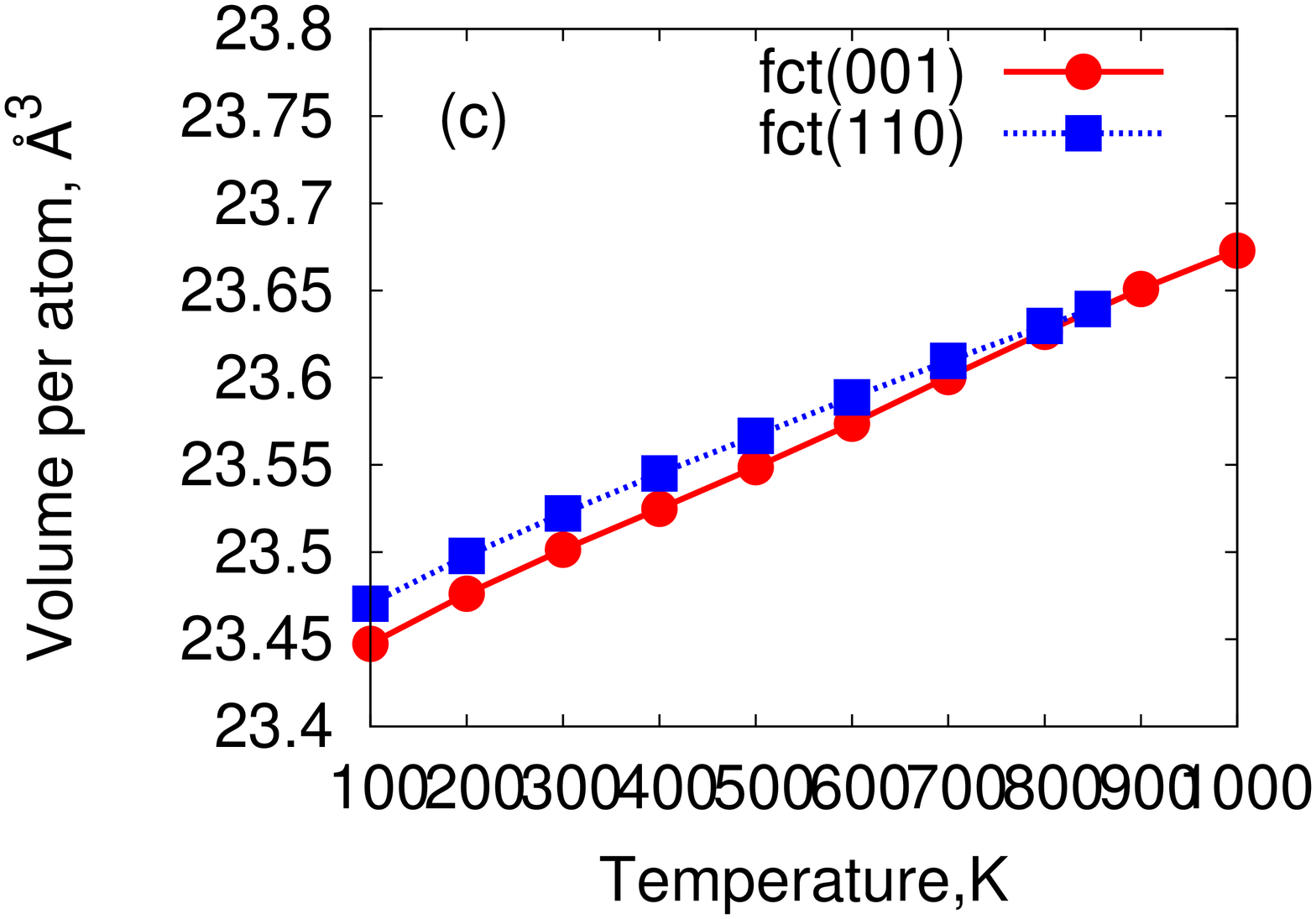}}
\resizebox{0.49\columnwidth}{!}{\includegraphics*[angle=-90]{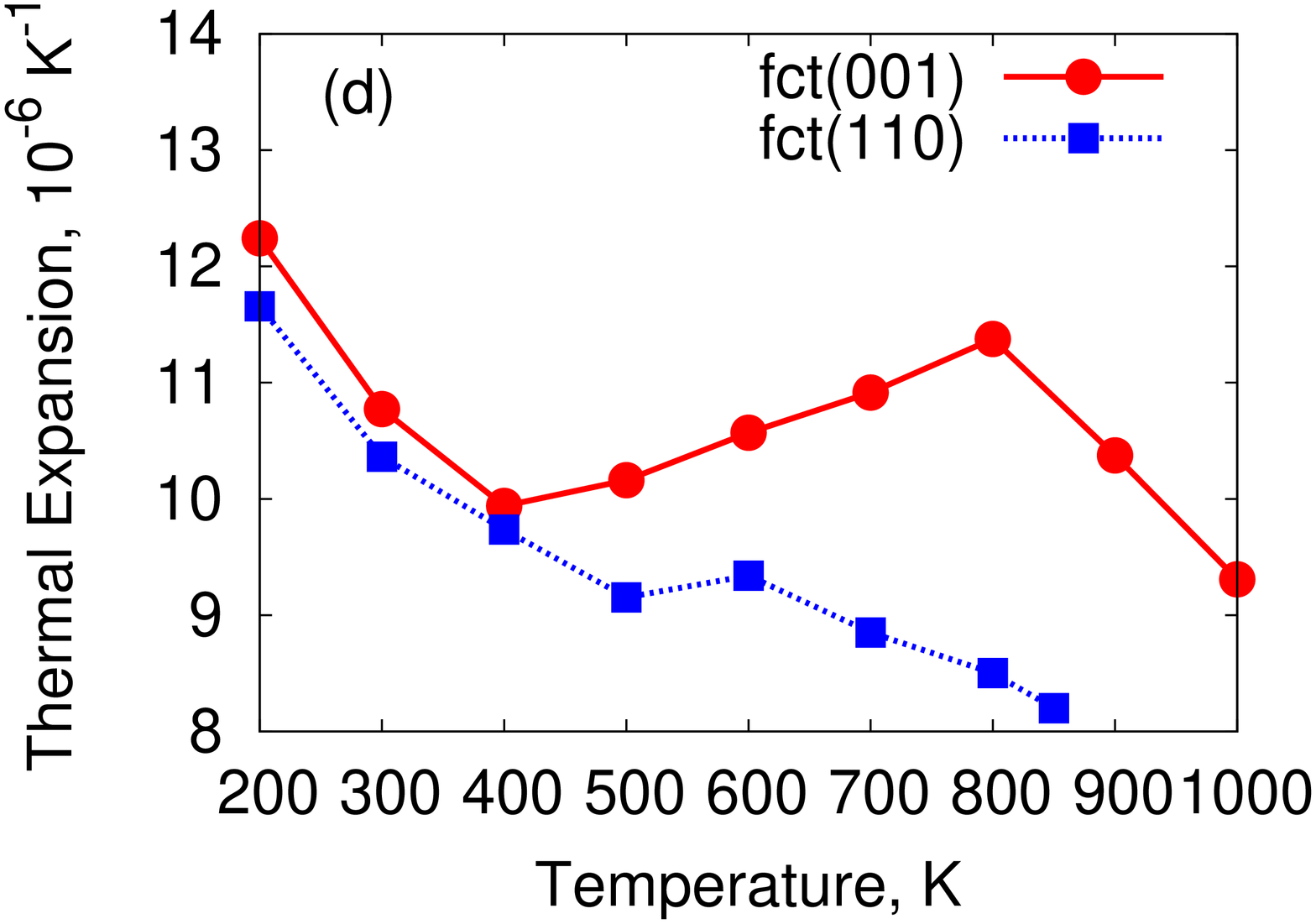}}
\caption{Temperature dependence of linear CTE for the films $fct$ (001)
(red) and $fct$ (110) (blue): along the normal to the film plane (a); in the
film plane (b). Temperature dependence of the volume per atom of $fct$ films
(c); temperature dependence of the CTE  for $fct$ films (d).}

\label{Fig8}
\end{center}
\end{figure}

The temperature dependences of the linear CTE were
also calculated for the $fct$ zirconium films. In Fig.\ref{Fig8} the linear CTE
are shown for the films $fct$ (001) (red) and $fct$ (110) (blue): $\alpha_{z}$
along the normal to the film plane (a); $\alpha_{x}$ and  $\alpha_{y}$ in the
film plane(b). 

The $fct$ (001) film has negative coefficients of linear expansion along the
axes parallel to the film plane over the whole temperature range considered. In
the $fct$ (110) film the linear CTE is negative only along the $x$ axis,
and along the $y$ direction it is positive. Note that the $fct$ (110) film
becomes unstable even at $T=900K$, there occurs a transition into an $hcp$
structure. The change of volume  per atom in the film interior is shown in
Fig.\ref{Fig8}(c) as a function of the temperature. In both films the volume
increases owing to a more rapid growth of the parameter $c$. The temperature
dependences of CTE are presented in Fig.\ref{Fig8}(d) for the $fct$ films (001)
(red) and (110) (blue).

Thus, calculations of the coefficients of thermal expansion and linear CTE for
zirconium films from the molecular dynamics simulation show that for the stable
$bct$ (110) film all CTE are positive throughout the temperature
interval investigated. The linear CTE of the $bct$(001) film in the film
plane are alternating-sign, the sign being changed to negative just
before the structural transformation. 

For metastable $fct$ (001) and (110) films CTE remain positive, but
their behavior depends substantially on the film surface orientation. At high
temperature the coefficient is of essentially nonlinear character for the
$fct$ (001) film, and also for the $fct$ (110) film the CTE curve exhibits a
hump. In the whole temperature range the linear CTE $\alpha_{x}$ and
$\alpha_{y}$ are negative for the $fct$ (001) film, whereas for $fct$ (110) only
the coefficient $\alpha_{x}$ is negative, and $\alpha_{y}$ is greater than zero.

\section{Conclusion}

The calculation of the total and local vibrational densities of  states
polarized along the axes $x, y, z$  as a function of temperature for the surface
and interior layers of $bct$ (stable) and $fct$ (metastable) zirconium films has
shown a correlation between the VDOS of surface atomic layers and the
anisotropy of the lattice parameters variation with temperature. For example, in
(100) and (110) $fct$ Zr films the decrease of the lattice parameters with
increasing temperature is due to the ``softening'' of low-frequency peaks in the
VDOS curves. Furthermore, it is shown that at temperatures close to
that of the structural transition the zirconium films can also exhibit 
negative linear coefficients of thermal expansion.

It is known that the negative CTE observed in bulk samples of composite
materials (ceramics, alloys) either are associated with vibrations of bridging
oxygen (Me-O-Me) [1], or they arise in highly anisotropic layered structures as
a result of the membrane effect [20]. In the case of zirconium thin films
neither mechanism of the appearance of negative linear CTE is feasible, firstly,
because of the lack of light atoms necessary for the bridging mechanism, and
secondly, the existing anisotropy of interaction between the atoms in the film
plane and in the direction normal to the plane does not suffice for the
membrane effect to arise.
%Возможно в ``мостиковом'' механизме необязательно должен быть легкий атом в
%связи Me-O-Me, а достаточно чтобы материал был сильноангармоничным и
%находился в метастабильном состоянии, что и приводит к появлению эффекта
%``смягчения'' низко-энергетических мод и уменьшению параметров решетки.
% Таким образом, в любом случае, причины появления ...
Thus, the reasons for the appearance of negative
coefficients of linear thermal expansion in zirconium films require further
investigation.

\section*{Acknowledgments}

The authors acknowledge  the partial support from the RFBR Grant $N^{o}$
13-02-96023-r-ural-a.

\end{document}